\documentclass[10pt, conference, compsocconf]{IEEEtran}

\usepackage{graphicx}
\usepackage{times}
\usepackage{helvet}
\usepackage{courier}
\usepackage{booktabs}
\usepackage[numbers,sort]{natbib}
\usepackage{epsfig}
\usepackage{amssymb}
\usepackage{amsmath}
\usepackage{amsfonts}
\usepackage{breqn}
\usepackage{multirow}
\usepackage{array}
\usepackage{subcaption}
\usepackage{mathtools}
\usepackage{caption}
\usepackage{url}
\usepackage{float}
\usepackage{balance}
\usepackage{algorithm}
\usepackage{algorithmic}
\usepackage{bbding}
\newcommand{\eq}[1]{Eq.~(\ref{#1})}
\newcommand{\xhdr}[1]{\vspace{0mm}\noindent{{\bf #1}}}
\newcommand{\md}{\emph{Fossil}}
\newcommand{\modelnamelong}{Factorized Sequential Prediction with Item Similarity Models}
\DeclareMathOperator*{\argmax}{arg\,max}
\DeclareMathOperator*{\argmin}{arg\,min}

\begin{document}

\title{Fusing Similarity Models with Markov Chains for Sparse Sequential Recommendation}

\author{\IEEEauthorblockN{Ruining He, Julian McAuley}
\IEEEauthorblockA{Department of Computer Science and Engineering\\
University of California, San Diego\\
Email: \{r4he, jmcauley\}@cs.ucsd.edu}
}

\maketitle
\begin{abstract}
Predicting personalized sequential behavior is a key task for recommender systems. In order to predict user actions such as the next product to purchase, movie to watch, or place to visit, it is essential to take into account both long-term user preferences and sequential patterns (i.e., short-term dynamics). Matrix Factorization and Markov Chain methods have emerged as two separate but powerful paradigms for modeling the two respectively. Combining these ideas has led to unified methods that accommodate long- and short-term dynamics simultaneously by modeling pairwise user-item and item-item interactions. 

In spite of the success of such methods for tackling dense data, they are challenged by \emph{sparsity} issues, which are prevalent in real-world datasets. In recent years, \emph{similarity}-based methods have been proposed for (sequentially-unaware) item recommendation with promising results on sparse datasets. In this paper, we propose to fuse such methods with Markov Chains to make personalized sequential recommendations. We evaluate our method, \md, on a variety of large, real-world datasets.
We show quantitatively that \md{} 
outperforms alternative algorithms, especially on sparse datasets, and qualitatively that it captures personalized dynamics and is able to make meaningful recommendations. 
\end{abstract}


\IEEEpeerreviewmaketitle

\section{Introduction}
Modeling and understanding the interactions between users and items, as well as the relationships amongst the items themselves are the core tasks of a recommender system. The former helps answer questions like `What kind of item does this specific user like?' (item-to-user recommendation), and the latter `Which type of shirts match the pants just purchased?' (item-to-item recommendation). In other words, (long-term) user preferences and (short-term) sequential patterns are captured by the above two forms of 
interactions 
respectively.

In this paper, we are interested in predicting personalized \emph{sequential} behavior from collaborative data (e.g.~purchase histories of users), which is challenging as long- and short-term dynamics need to be combined carefully to account for both personalization and sequential transitions. This challenge is further complicated by \emph{sparsity} issues in many real-world datasets, which makes it hard to estimate parameters accurately from limited training sequences. Particularly, this challenge is not addressed by models 
concerned with historical temporal dynamics (e.g.~the popularity fluctuation of \emph{Harry Potter} between 2002 to 2006), where user-level sequential patterns are typically ignored (e.g.~`what will Tom watch next after watching \emph{Harry Potter}?').

\begin{figure}[!t]
\centering
\includegraphics[width=\linewidth]{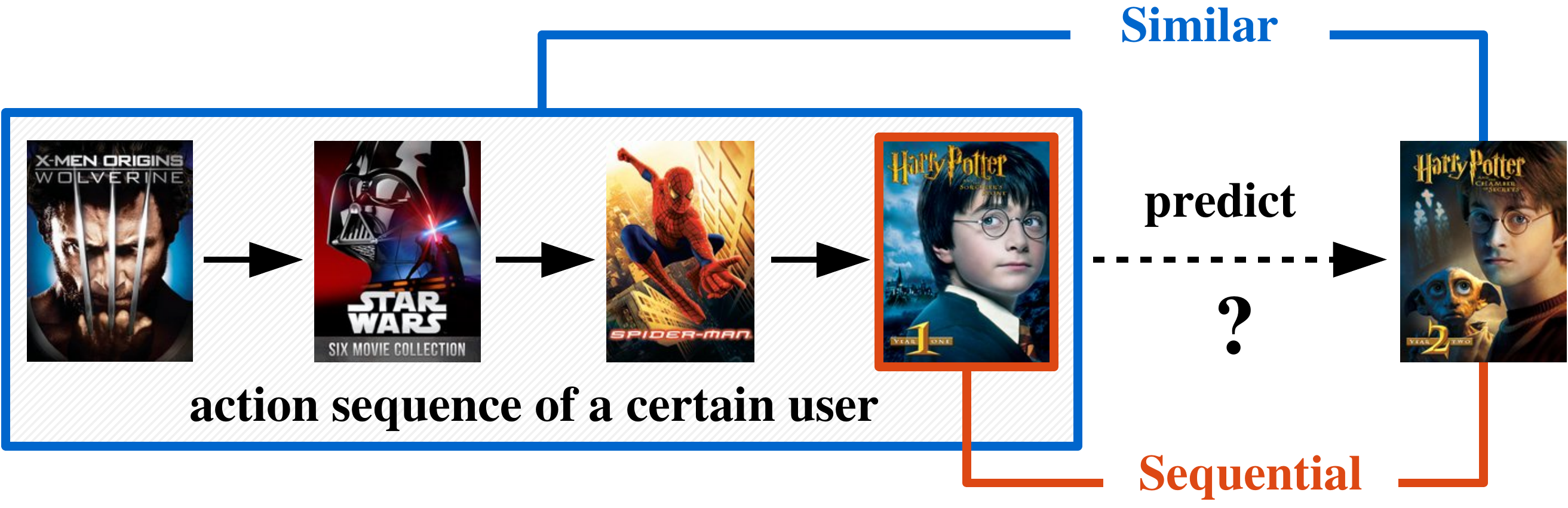}
\caption{An example of how our method, \md, makes recommendations. \emph{Harry Potter 2} is recommended to the user because it (1) is similar to content already watched (i.e., fantasy movies), and (2) frequently follows the recently-watched movie \emph{Harry Potter 1}. The former is modeled with a similarity-based method and the latter Markov Chains.}
\label{fig:idea}
\end{figure}

To model user preferences, there have been two relevant streams of work. Traditional item recommendation algorithms are typically based on a low-rank factorization of the user-item interaction matrix, referred to as Matrix Factorization \cite{Handbook}. Each user or item is represented with a numerical vector of the same dimension such that the compatibility between them is estimated by the inner product of their respective representations. Recently, an \emph{item similarity}-based algorithm---\emph{Factored Item Similarity Models} (FISM)---has been developed which makes recommendations to a user $u$ exclusively based on how similar items are to those already consumed/liked by $u$. In spite of not explicitly parameterizing each user, FISM surprisingly outperforms various competing baselines, including Matrix Factorization, especially on sparse datasets \cite{kabbur2013fism}.

The above methods are unaware of sequential dynamics. In order to tackle sequential prediction tasks, we need to resort to alternate methods, such as Markov Chains, that are able to capture sequential patterns. To this end, Rendle \emph{et al.} proposed \emph{Factorized Personalized Markov Chains} (FPMC), which combines Matrix Factorization and a first-order Markov Chain \cite{rendle2010fpmc} to model personalized sequential behavior.

Despite the success achieved by FPMC, it suffers from {sparsity} issues and the long-tailed distribution of many datasets, so that the sequential prediction task is only {partially} solved. In this paper, we propose to fill this gap by fusing similarity-based methods (like FISM) with Markov Chain methods (like FPMC) to tackle sparse real-world datasets with sequential dynamics. The resulting method, \textbf{F}act\textbf{O}rized \textbf{S}equential Prediction with Item \textbf{SI}milarity Mode\textbf{L}s (or \md{} in short), naturally combines the two by learning a personalized weighting scheme over the sequence of items to characterize users in terms of both preferences and the strength of sequential behavior. Figure \ref{fig:idea} demonstrates an example of how \md{} makes recommendations. 

\md{} brings the following benefits for tackling sparsity issues: (1) It parameterizes each user with only the historical items so that cold-user (or `cool-user') issues can be alleviated so long as the representations of items can be estimated accurately. (2) For cold-users, \md{} can shift more weight to short-term dynamics to capitalize from `global' sequential patterns. This flexibility enables \md{} to make reasonable predictions even though only a few actions may have been observed for a given user.

Our contributions are summarized as follows: First, we develop a new method, \md, that integrates similarity-based methods 
with
Markov Chains smoothly to make personalized sequential predictions on sparse and long-tailed datasets. Second, we demonstrate quantitatively that \md{} is able to outperform a spectrum of state-of-the-art algorithms on a variety of large, real-world datasets with around five million user actions in total. Finally, we visualize the learned model and analyze the sequential and personalized dynamics captured.

Data and code are available at \url{https://sites.google.com/a/eng.ucsd.edu/ruining-he/}.

\section{Related Work}
The most closely related works to ours are (1) Item recommendation methods that model user preferences but are \emph{unaware} of sequential dynamics; (2) Works that deal with temporal dynamics but rely on explicit time stamps; and (3) Those that address the sequential prediction task we are interested in.

\xhdr{Item recommendation.} Item recommendation usually relies on Collaborative Filtering (CF) to learn from explicit feedback like star-ratings \cite{Handbook}. CF predicts based only on the user-item rating matrix and mainly follows two paradigms: neighborhood- and model-based. 
Neighborhood-based methods 
recommend
items that either have been enjoyed by like-minded users (user-oriented, e.g.~\cite{konstan1997grouplens,shardanand1995social,hill1995recommending}) or are similar to those already consumed (item-oriented, e.g.~\cite{AmazonItemToItem, deshpande2004item}). Such methods have to rely on some type of \emph{predefined} similarity metric such as {Pearson Correlation} or {Cosine Similarity}. 
In contrast, model-based methods directly explain the interactions between users and items. There have been a variety of such algorithms including Bayesian methods \cite{miyahara2000collaborative, breese1998empirical}, Restricted Boltzmann Machines \cite{salakhutdinov2007restricted}, Matrix Factorization (MF) methods (the basis of many state-of-the-art recommendation approaches such as \cite{BellKorSolution}, \cite{Netflixprize}, \cite{NSVD}), and so on. 

In order to tackle implicit feedback data where only positive signals (e.g.~purchases, clicks, thumbs-up) are observed, both neighborhood- and model-based methods have been extended. Recently, Ning \emph{et al.}~proposed SLIM to learn an item-item similarity matrix, which has shown to outperform a series of state-of-the-art recommendation approaches \cite{ning2011slim}. Kabbur \emph{et al.}~further explored the low-rank property of the similarity matrix to handle sparse datasets \cite{kabbur2013fism}. Since similarity (or neighborhood) relationships are learned from the data, these methods overcome the rigidity of using a predefined similarity metric. On the other hand, MF has also been extended in several ways including
point-wise methods that inherently assume non-observed feedback to be negative \cite{WRMF,OCCF}, and pair-wise methods like BPR-MF \cite{rendle2009bpr} that are based on a more realistic assumption that positive feedback should only be `more preferable' than non-observed feedback. 

\xhdr{Temporal dynamics.} 
Several works take temporal dynamics into account, mostly based on MF techniques \cite{koren2008factorization}. 
This includes seminal
work proposed by Koren~\cite{koren2010temporal,koren2009matrix}, where they showed state-of-the-art results on \emph{Netflix} data by modeling the evolution of users and items over time.
However, such works are ultimately building models to understand 
past actions
(e.g.~`What did Tom like in 2008?', `What does Grace like to do on Weekends?'), by making use of the explicit time stamps. The sequential prediction task differs from theirs in that it does not use time stamps directly, but rather models sequential relationships between actions.

\xhdr{Sequential recommendation.} 
Markov Chains have demonstrated their strength at modeling stochastic transitions, from uncovering sequential patterns (e.g.~\cite{zimdars2001using,mobasher2002using}) to directly modeling decision processes \cite{shani2002mdp}.
For the sequential prediction/recommendation task, Rendle \emph{et al.}~proposed FPMC which combines the power of MF at modeling {personal} preferences and the strength of Markov Chains at modeling sequential patterns \cite{rendle2010fpmc}. 
Our work follows this thread but contributes in that (1) we make use of a similarity-based method for modeling user preferences so that sparsity issues are mitigated; and (2) we further consider Markov Chains with higher orders to 
model sequential smoothness across multiple time steps.

\section{Sequential Prediction} 

\subsection{Problem Formulation and Notation}  \label{sec:formulate}
Objects to be recommended in the system are referred to as \emph{items}. The most common recommendation approaches focus on modeling types of items of interest to each user, without accounting for any \emph{sequential} information, e.g., the last item purchased/reviewed, or place visited by the user in question. 

We are tackling \emph{sequential prediction} tasks which are formulated as follows. Let $\mathcal{U} = \{u_1, u_2, \ldots, u_{|\mathcal{U}|}\}$ denote the set of users and $\mathcal{I} = \{i_1, i_2, \ldots, i_{|\mathcal{I}|}\}$ the set of items.
Each user $u$ is associated with a \emph{sequence} of actions (or `events') $\mathcal{S}_u$ (e.g.~items purchased by $u$, or places $u$ has checked in): $\mathcal{S}^u = (\mathcal{S}^u_1, \mathcal{S}^u_2, \ldots, \mathcal{S}^u_{|\mathcal{S}^u|})$, where $\mathcal{S}^u_k \in \mathcal{I}$. We use $\mathcal{I}^+_u$ to denote the \emph{set} of items in $\mathcal{S}^u$ where the sequential signal is ignored. 
Using the above data exclusively, our objective is to predict the next action of each user and thus make recommendations accordingly.
Notation used throughout this paper is summarized in Table \ref{tab:notation}.
Boldfaced symbols are used to denote matrices.

\begin{table}
\centering
\setlength{\tabcolsep}{4pt}
\caption{Notation \label{tab:notation}}
\begin{tabular}{lp{0.8\linewidth}} \toprule
Notation & Explanation\\ \midrule
$\mathcal{U}, \mathcal{I}$ & user set, item set\\
$u, i, t$ & a specific user, item, time step \\
$\mathcal{S}^u_t$ & the item user $u$ interacted with at time step $t$\\
$\mathcal{S}^u$ & action sequence of user $u$, $\mathcal{S}^u = (\mathcal{S}^u_1, \mathcal{S}^u_2, \ldots, \mathcal{S}^u_{|\mathcal{S}^u|})$ \\
$\mathcal{I}^+_u$ & the set of items in $\mathcal{S}^u$, $\mathcal{I}^+_u = \{\mathcal{S}^u_1, \mathcal{S}^u_2, \ldots, \mathcal{S}^u_{|\mathcal{S}^u|}\}$ \\
$\beta_i$ & bias term associated with item $i$, $\beta_i \in \mathbb{R}$ \\
$\mathbf{P}_i$ & latent vector associated with item $i$, $\mathbf{P}_i \in \mathbb{R}^K$ \\
$\mathbf{Q}_i$ & latent vector associated with item $i$, $\mathbf{Q}_i \in \mathbb{R}^K$ \\
$K$ & dimensionality of the vector representing each user/item \\
$L$ & order of Markov Chains \\
$\eta$ & global weighting vector, $\eta = (\eta_1, \eta_2, \ldots, \eta_L)$ \\
$\eta^u$ & personalized weighting vector, $\eta^u = (\eta^u_1, \eta^u_2, \ldots, \eta^u_L)$ \\
$p_u(j~|~i)$ & probability that user $u$ chooses item $j$ after item $i$ \\ 
$\widehat{p}_{u,t,i}$ & prediction that user $u$ chooses item $i$ at time step $t$ \\ 
$>_{u,t}$ & personalized total order of user $u$ at time step $t$ \\
$\alpha$ & weighting factor, $\alpha \in \mathbb{R}$ \\ 
$\epsilon$ & learning rate, $\epsilon \in \mathbb{R}$ \\ 
$\sigma(\cdot)$ & the logistic (sigmoid) function \\ 
\bottomrule
\end{tabular}
\end{table}

\subsection{Modeling User Preferences} \label{sec:longterm}
Modeling and understanding long-term preferences of users is a key to any recommender system. Traditional methods such as Matrix Factorization (e.g.~\cite{korenSurvey}) are usually based on a low-rank assumption. They project users and items to a low-rank latent space ($K$-dimensional) such that the coordinates of each user within the space capture the \emph{preferences} towards these $K$ latent dimensions. The affinity $\widehat{p}_{u,i}$ between user $u$ and item $i$ is then estimated by the inner product of the vector representations of $u$ and $i$:
\begin{equation} \label{eq:mf}
\widehat{p}_{u,i} = \langle \mathbf{X}_u, \mathbf{Y}_i \rangle.
\end{equation}

Recently, a novel similarity-based method, called \emph{Sparse Linear Methods} (SLIM), has been developed and shown to outperform a series of state-of-the-art approaches including Matrix Factorization based methods \cite{ning2011slim}. By learning an item-to-item similarity matrix $\mathbf{W}$ from the user action history (e.g.~purchase logs), it predicts the user-item affinity as follows:
\begin{equation}
\widehat{p}_{u,i} = \sum_{j \in \mathcal{I}^+_u \setminus \{i\}} \mathbf{W}_{j,i},
\end{equation}
where $\mathcal{I}^+_u$ is the set of items $u$ has interacted with. $\mathbf{W}_{j,i}$ is the element at the $j$-th row and $i$-th column of matrix $\mathbf{W}$, denoting the \emph{similarity} of item $j$ to item $i$. The underlying rationale it follows is that the more $j$ is similar to those items already consumed/liked by $u$, the more likely $j$ will be a preferable choice for $u$.  

Without parameterizing each user explicitly, SLIM relaxes the low-rank assumption enforced on user representations and has achieved higher recommendation accuracy (see \cite{ning2011slim} for details).

The major challenge faced by SLIM comes from the large amount of parameters ($|\mathcal{I}| \times |\mathcal{I}|$) to be estimated from the sparse user-item interactions. SLIM approaches this issue by exploring the \emph{sparsity} characteristic of $\mathbf{W}$ using $\mathcal{L}_1$-norm regularization when inferring the parameters. Another direction is to capitalize on the low-rank potential of the similarity matrix by decomposing $\mathbf{W}$ into the product of two independent low-rank matrices \cite{kabbur2013fism}:
\begin{equation} \label{eq:wpq}
\mathbf{W} = \mathbf{P} \mathbf{Q}^T, 
\end{equation}
where $\mathbf{P}$ and $\mathbf{Q}$ are both $|\mathcal{I}| \times K$ matrices and $K \ll |\mathcal{I}|$. This method is called \emph{Factored Item Similarity Models} (FISM) and brings two benefits: (1) It significantly reduces the number of parameters and has been shown to generate state-of-the-art performance on a series of \emph{sparse} datasets; and (2) Compared to SLIM, it is stronger at capturing the \emph{transitive} property of item similarities.\footnote{For instance, if items $a$ and $b$, $b$ and $c$ are two co-purchase pairs in the training data but $(a, c)$ is not, SLIM will erroneously estimate the similarity of $a$ and $c$ to be $0$.} When it comes to real-world datasets which are usually highly sparse, the above benefits contribute considerably to recommendation performance.

\subsection{Modeling Sequential Patterns} \label{sec:temp}
Sequential (or short-term) dynamics are typically modeled by Markov Chains. Given the last item $i$ that has been interacted with, the first-order Markov Chain predicts the probability of item $j$ being chosen at the next step $p(j~|~i)$ by maximum likelihood estimation of the item-to-item transition matrix. A further improvement can be made by factorizing the transition matrix into two low-rank matrices, similar to the idea in Section \ref{sec:longterm} (i.e., \eq{eq:wpq}); 
thus the transition probability of item $i$ to item $j$ is estimated by the following inner product: 
\begin{equation} \label{eq:mc}
p(j~|~i) \propto \langle \mathbf{M}_i, \mathbf{N}_j \rangle,
\end{equation}
where $\mathbf{M}_i$ and $\mathbf{N}_j$ are the latent vector representations of item $i$ and $j$ respectively. Note that each item $i$ is associated with two vectors: $\mathbf{M}_i$ and $\mathbf{N}_i$.

Markov Chains are strong at capturing short-term dynamics. For instance, if one purchased a laptop recently, it is reasonable to recommend relevant items such as peripherals or laptop backpacks. Nevertheless, such methods are limited in their ability to capture user preferences that are both personal and long-term 
(e.g.~what type of laptop backpack this particular user likes). 
Thus there is a need to combine the two lines of models carefully in order to benefit from modeling long- and short-term dynamics simultaneously.

\subsection{The Unified Sequential Prediction Model}
Recently, Rendle \emph{et al.} introduced a seminal method that combines Matrix Factorization (i.e., \eq{eq:mf}) and the first-order Markov Chain (i.e., \eq{eq:mc}) to form a unified prediction model, which is dubbed \emph{Factorized Personalized Markov Chains} (FPMC) \cite{rendle2010fpmc}. The probability that an arbitrary user $u$ transitions from the last item $i$ to the next item $j$ is estimated by
\begin{equation} \label{eq:fpmc}
p_u(j~|~i) \propto \langle \mathbf{X}_u, \mathbf{Y}_j \rangle + \langle \mathbf{M}_i, \mathbf{N}_j \rangle,
\end{equation}
where the first inner product computes how much $u$ likes item $j$ and the second calculates the extent to which $j$ is `similar' to the last item $i$.\footnote{In \cite{rendle2010fpmc}, the authors took a tensor-factorization perspective of the predictor, which brings an additional term modeling the interactions between $u$ and the last item $i$.
However, this term is not required as it always gets canceled out when making predictions.
}

\section{The Proposed \md{} Model} \label{sec:model}
\subsection{The Basic Model}
\begin{figure}
\centering
\includegraphics[width=\linewidth]{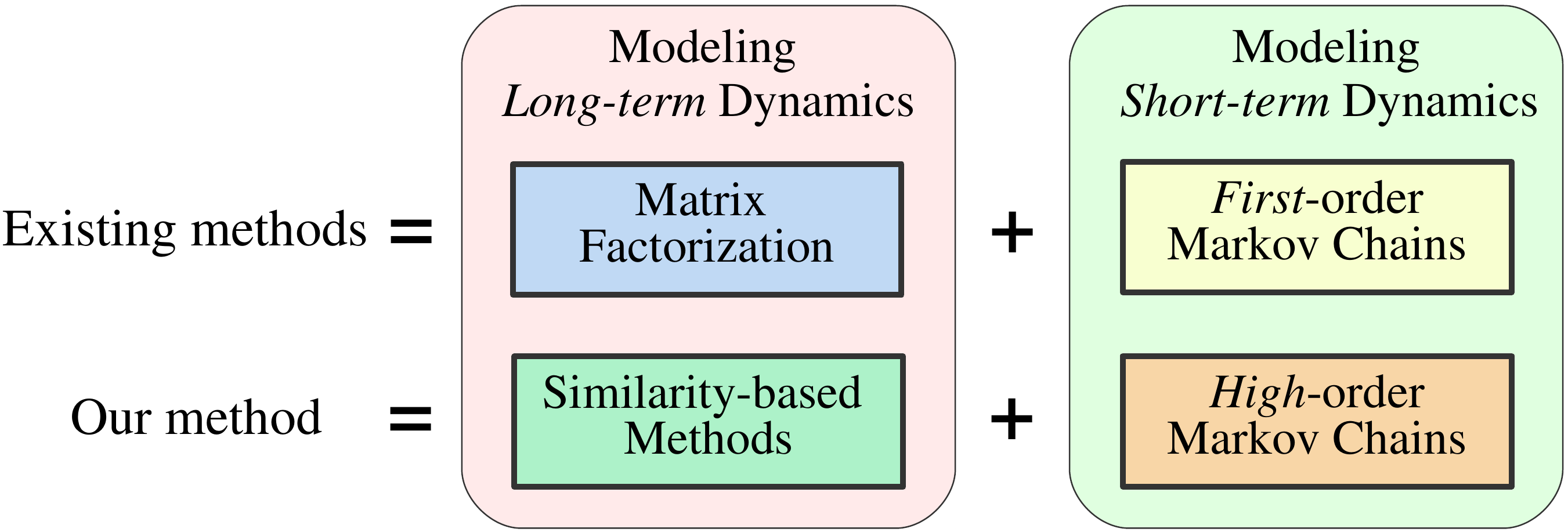}
\caption{Our method, \md, is a combination of similarity-based methods and (high-order) Markov Chains, in contrast to existing methods that combine Matrix Factorization and first-order Markov Chains \cite{rendle2010fpmc}.}
\label{fig:idea2}
\end{figure}

In contrast to FPMC,
here we take another direction and investigate combining similarity-based methods and Markov Chains to approach the sequential prediction task (see Figure \ref{fig:idea2}). In particular, we take FISM (see Section \ref{sec:longterm}) as our starting point, 
in light of its ability to handle the sparsity issues in real-world datasets.

The basic form of our model is as follows:
\begin{equation} \label{eq:our}
p_u(j~|~i) \propto \overbrace{\sum_{j' \in \mathcal{I}^+_u \setminus \{j\}} \langle \mathbf{P}_{j'}, \mathbf{Q}_j \rangle}^{\text{user preferences}} + \underbrace{(\eta + \eta_u)}_{\mathclap{\substack{\text{personalized} \\\ \text{weighting factor}}}} \cdot \overbrace{\langle \mathbf{M}_i, \mathbf{N}_j \rangle}^{\mathclap{\text{sequential dynamics}}},
\end{equation}
where each user is parameterized with only a single scalar $\eta_u$ that controls the relative weights of the long- and short-term dynamics. $\eta$ is a global parameter shared by all users and helps center $\eta_u$ at $0$.

\begin{figure*}
\begin{center}
\begin{equation} \label{eq:highorder}
 p_u(j~|~\underbrace{\mathcal{S}^u_{t-1}, \mathcal{S}^u_{t-2},\ldots, \mathcal{S}^u_{t-L}}_{\mathclap{\text{previous $L$ items $u$ interacted with}}})~~ \propto ~~ \beta_j + \langle \overbrace{\frac{1}{|\mathcal{I}^+_u \setminus \{j\}|^\alpha} \sum_{j' \in \mathcal{I}^+_u \setminus \{j\}} \mathbf{P}_{j'}}^{\text{long-term dynamics}} + \underbrace{\sum^L_{k=1}(\eta_k + \eta^u_k) \cdot \mathbf{P}_{\mathcal{S}^u_{t-k}}}_{\text{short-term dynamics}}, \mathbf{Q}_j \rangle.
\end{equation}
\end{center}
\end{figure*}

The above formulation parameterizes each item with four vectors, i.e., $\mathbf{P}_i$, $\mathbf{Q}_i$, $\mathbf{M}_i$, and $\mathbf{N}_i$. Considering the limited number of parameters we can afford in the sparse datasets we are interested in, we reduce the four matrices to two by enforcing $\mathbf{P} = \mathbf{M}$ and $\mathbf{Q} = \mathbf{N}$. 
This makes sense since ultimately sequentially-related items are also  `similar' to one another. Adding a bias term $\beta_j$ and normalizing the long-term dynamics component,
we arrive at a new formulation as follows:
\begin{equation} \label{eq:impr}
\begin{split}
p_u(j~|~i) &\propto  \beta_j + \langle \frac{1}{|\mathcal{I}^+_u \setminus \{j\}|^\alpha} \sum_{\mathclap{~~j' \in \mathcal{I}^+_u \setminus \{j\}}} \mathbf{P}_{j'} + (\eta + \eta_u) \cdot \mathbf{P}_i, \mathbf{Q}_j \rangle.
\end{split}
\end{equation}

\subsection{Modeling Higher-order Markov Chains}
Up to now we have used first-order Markov Chains to model short-term temporal dynamics. Next, we extend our formulation to consider high-order Markov Chains to capture smoothness across multiple time steps. Given the most recent $L$ items user $u$ has consumed ($\mathcal{S}^u_{t-1}, \mathcal{S}^u_{t-2},\ldots, \mathcal{S}^u_{t-L}$), the new formulation predicts the probability of item $j$ being the next item (at time step $t$) with an $L^{\text{th}}$ order Markov Chain as shown in \eq{eq:highorder}. In this new formulation, each user is associated with a vector $\eta^u = (\eta^u_1, \eta^u_2, \ldots, \eta^u_L)$. Likewise, the global bias becomes the vector $\eta = (\eta_1, \eta_2, \ldots, \eta_L)$. The rationale behind this idea is that each of the previous $L$ actions should contribute with different weights to the high-order smoothness.

\subsection{Inferring Model Parameters}
The ultimate goal of the sequential prediction task is to rank observed (or ground-truth) items as high as possible so that the recommender system can make plausible recommendations. This means it is natural to derive a \emph{personalized} total order $>_{u,t}$ 
(at each step $t$) to minimize a ranking loss such as sequential Bayesian Personalized Ranking (S-BPR) \cite{rendle2010fpmc}. Here $i >_{u,t} j$ means that item $i$ is ranked higher than item $j$ for user $u$ at step $t$ given the action sequence before $t$. 

For each user $u$ and for each time step $t$, S-BPR employs a sigmoid function $\sigma(\widehat{p}_{u,t,\mathcal{S}^u_t} - \widehat{p}_{u,t,j})$ ($\widehat{p}_{u,t,\cdot}$ is a shorthand for the prediction in \eq{eq:highorder}) to characterize the probability that ground-truth item $\mathcal{S}^u_t$ is ranked higher than a `negative' item $j$ given the model parameters $\Theta$, i.e., $p(\mathcal{S}^u_t >_{u,t} j | \Theta)$. 
Assuming independence of users and time steps, model parameters $\Theta$ are inferred by optimizing the following \emph{maximum a posteriori} (MAP) estimation:
\begin{equation}
\begin{split}
\argmax_{\Theta} = \ln \prod_{u \in \mathcal{U}} \prod_{t = 2}^{|\mathcal{S}^u|} \prod_{j \neq \mathcal{S}^u_t} p(\mathcal{S}^u_t >_{u,t} j | \Theta) ~ p(\Theta) \\
= \sum_{u \in \mathcal{U}} \sum_{t = 2}^{|\mathcal{S}^u|} \sum_{j \neq \mathcal{S}^u_t} \ln \sigma(\widehat{p}_{u,t,\mathcal{S}^u_t} - \widehat{p}_{u,t,j}) + \ln p(\Theta),
\end{split}
\end{equation}
where the \emph{pairwise} ranking between the ground-truth and all negative items goes through all users and all time steps.\footnote{i.e., optimize the AUC metric (See Section \ref{sec:metric}).} $p(\Theta)$ is a Gaussian prior over the model parameters. Note that due to the `additive' characteristic, our formulation can allow $t$ to run from $2$ (instead of $L+1$) to the last item in $\mathcal{S}^u$.

The large amount of positive-negative pairs in the objective makes conventional batch gradient descent unaffordable. As such, we adopt Stochastic Gradient Descent (SGD) which has seen wide success for learning models in BPR-like optimization frameworks (e.g.~\cite{rendle2009bpr,rendle2010fpmc}). The SGD training procedure works as follows. First, it uniformly samples a user $u$ from $\mathcal{U}$ as well as a time step $t$ from $\{2,3,\ldots,|\mathcal{S}^u|\}$. Next, a negative item $j \in \mathcal{I}$ and $j \notin \{\mathcal{S}^u_t, \mathcal{S}^u_{t-1}, \ldots, \mathcal{S}^u_{t-\min(L,t-1)}\}$ is uniformly sampled, which forms a training triple $(u,t,j)$. Finally, the optimization procedure updates parameters in the following fashion: 
\begin{equation} \label{eq:sgd}
\Theta \leftarrow \Theta + \epsilon \cdot (\sigma(\widehat{p}_{u,t,j}-\widehat{p}_{u,t,\mathcal{S}^u_t}) ~ \frac{\partial (\widehat{p}_{u,t,\mathcal{S}^u_t} - \widehat{p}_{u,t,j})}{\partial \Theta} - \lambda_{\Theta}\Theta ), 
\end{equation}
where $\epsilon$ is the learning rate and $\lambda_{\Theta}$ is a regularization hyperparameter. 

\section{Experiments}
\subsection{Datasets and Statistics} \label{sec:dataset}
To evaluate the ability
and applicability 
of our method to handle different real-world scenarios, we include a spectrum of large datasets from different domains in order to predict actions ranging from the next product to purchase, next movie to watch, to next review to write, and next place to check-in. Note that these datasets also vary significantly in terms of user and item density (i.e., number of actions per user/item). 

\xhdr{\emph{Amazon}.} The first group of large datasets are from \emph{Amazon.com},\footnote{https://www.amazon.com/} recently introduced by \cite{VisualSIGIR}.
This is among the largest datasets available that includes
review texts and time stamps spanning from May 1996 to July 2014. Each top-level category of products on \emph{Amazon.com} has been constructed as an independent dataset by the authors of \cite{VisualSIGIR}. In this paper, we take a variety of large categories including office products, automotive, video games, toys and games, cell phones and accessories, clothing, shoes, and jewelry, and electronics.

\xhdr{\emph{Epinions}.} The next dataset is collected from a popular online consumer review website \emph{Epinions.com}\footnote{http://epinions.com/} by the authors of \cite{zhao2014leveraging}. Like \emph{Amazon} data, it includes all actions of all users on the website so that the sequential relationships are maintained. This dataset spans January 2001 to November 2013. 

\begin{table}
\begin{center}
\setlength{\tabcolsep}{3pt}
\caption{Statistics of the Datasets.}\label{table:data}
\begin{tabular}{lrrrcc} \toprule
Dataset    			&\parbox{0.12\linewidth}{\centering\#users ($|\mathcal{U}|$)}    &\parbox{0.12\linewidth}{\centering\#items ($|\mathcal{I}|$)}  &\#actions   &\parbox{0.12\linewidth}{\centering avg. \#actions /user}  &\parbox{0.12\linewidth}{\centering avg. \#actions /item}  \\ \midrule 
\emph{Amazon-Office}&16,716      &22,357    &128,070     &7.66   &5.73  \\ 
\emph{Amazon-Auto}  &34,316      &40,287    &183,573     &5.35   &4.56  \\  
\emph{Amazon-Game}  &31,013      &23,715    &287,107     &9.26   &12.11 \\ 
\emph{Amazon-Toy}   &57,617      &69,147    &410,920     &7.13   &5.94  \\
\emph{Amazon-Cell}  &68,330      &60,083    &429,231     &6.28   &7.14  \\
\emph{Amazon-Clothes} &184,050     &174,484   &1,068,972   &5.81   &6.13  \\
\emph{Amazon-Elec}  &253,996     &145,199   &2,109,879   &8.31   &14.53 \\
\emph{Epinions}     &5,015       &8,335     &26,932      &5.37   &3.23  \\ 
\emph{Foursquare}   &43,110      &13,335    &306,553     &7.11   &22.99 \\ 
\textbf{Total}      &\textbf{694,163} &\textbf{556,942} &\textbf{4,951,237} &{N/A}    &{N/A}   \\ \bottomrule
\end{tabular}
\end{center}
\setlength{\tabcolsep}{4.2pt}
\raggedright{\textbf{\emph{Amazon} datasets are ranked by the `\#actions' column. \emph{Office}: Office Products, \emph{Auto}: Automotive, \emph{Game}: Video Games, \emph{Toy}: Toys and Games, \emph{Cell}: Cell Phones and Accessories, \emph{Clothes}: Clothing, Shoes and Jewelry, \emph{Elec}: Electronics.}}
\end{table}

\xhdr{\emph{Foursquare}.} We also include another popular dataset which is often used to evaluate next Point-Of-Interest prediction algorithms. It is from \emph{Foursquare.com}\footnote{https://foursquare.com/} and includes a large number of check-ins (i.e., visits) of users at different venues (e.g.~restaurants), spanning December 2011 to April 2012. Note that the setting in this paper is to compare all methods using collaborative data exclusively, as in Section \ref{sec:formulate}; making use of side signals like geographical data is beyond the scope of this work.  

For each of the above datasets, we filter out inactive users and items with fewer than five associated actions. Star-ratings are converted to implicit feedback (i.e., `binary' actions) by setting the corresponding entries to 1; that is, we care about the purchase/review/check-in actions regardless of the specific rating values. Statistics of each dataset after the above processing are shown in Table \ref{table:data}.

\subsection{Evaluation Methodology} \label{sec:metric}
All methods are evaluated with the AUC (\emph{Area Under the ROC curve}) metric, not only because it is widely used (e.g.~\cite{VBPR,rendle2009bpr}), but also because it is a natural choice in our case as all comparison methods directly optimize this metric on the training set. 

For each dataset, we use the two most recent actions of each user to create a validation set $\mathcal{V}$ and a test set $\mathcal{E}$: one action for validation and the other for testing.
All other actions are used as the training set $\mathcal{T}$. 
The training set $\mathcal{T}$ is used to train all comparison methods, and hyperparameters are tuned with the validation set $\mathcal{V}$. Finally, all trained models are evaluated on the test set $\mathcal{E}$:
\begin{equation}
\mathit{AUC} =  \frac{1}{|\mathcal{U}|}  \sum_{u \in \mathcal{U}}   \frac{1}{|\mathcal{I}\setminus \mathcal{I}^+_u|}   \sum_{j \in \mathcal{I} \setminus \mathcal{I}^+_u}  \mathbf{1} (\widehat{p}_{u,|\mathcal{S}^u|,g_u} > \widehat{p}_{u,|\mathcal{S}^u|,j}),
\end{equation}
where $g_u$ is the ground-truth item of user $u$ at the most recent time step $|\mathcal{S}^u|$. The indicator function $\mathbf{1}(b)$ returns $1$ if the argument $b$ is $\mathit{true}$, $0$ otherwise. 
The goal here is for the held-out action to calculate how highly the ground-truth item has been ranked for each user $u$ according to the learned personalized total order $>_{u,t}$.

\subsection{Comparison Methods}
We include a series of state-of-the-art methods in the field of both item recommendation and sequential prediction. 
\begin{enumerate}
\item{\textbf{Popularity (POP)}:} always recommends items based on the rank of their popularity in the system.

\item{\textbf{Bayesian Personalized Ranking (BPR-MF)} \cite{rendle2009bpr}:} is a state-of-the-art method for personalized item recommendation. It only considers long-term preferences and uses Matrix Factorization \cite{korenSurvey} as the underlying predictor.

\item{\textbf{Factored Item Similarity Models (FISM)} \cite{kabbur2013fism}:} is a 
recently-proposed similarity-based algorithm for personalized item recommendation. We build our model on top of it to tackle the sequential prediction task.

\item{\textbf{Factorized Markov Chains (FMC)}:} factorizes the item-to-item transition matrix ($|\mathcal{I}| \times |\mathcal{I}|$) to capture the likelihood that an arbitrary user transitions from one item to another. Here we use a first-order Markov Chain as higher orders incur a state-space explosion.

\item{\textbf{Factorized Personalized Markov Chain (FPMC)} \cite{rendle2010fpmc}:} is a 
method that uses a personalized Markov Chain (see \eq{eq:fpmc}) for the sequential prediction task we are interested in. Recall that FPMC is ultimately a combination of Matrix Factorization and first-order Markov Chains.

\item{\textbf{\modelnamelong{} (\md)}:} is the algorithm proposed in this paper (see \eq{eq:highorder}). Markov Chains of different orders will be experimented with and compared against other methods. 
\end{enumerate}

For clarity, the above methods are collated in Table \ref{table:base} in terms of whether they are `personalized,' `sequentially-aware,' `similarity-based,' `explicitly model users,' and `consider high-order Markov Chains.' 
Note that all methods (except POP) directly optimize the pairwise ranking of ground-truth actions versus negative actions in the training set $\mathcal{T}$ (i.e., the AUC metric), so that the fairness of the comparison is maximized.

\begin{table}
\begin{center}
\setlength{\tabcolsep}{6pt}
\caption{Models. {P: Personalized?, Q: seQuentially-aware?, S: Similarity-based?, E: Explicitly model users?, H: consider High-order Markov Chains?.}}\label{table:base}
\begin{tabular}{cccccccccc} \toprule
Property    &POP             &BPR-MF            &FISM           &FMC            &FPMC            &\md{}            \\ \midrule 
{P}  &\XSolidBrush    &\CheckmarkBold    &\CheckmarkBold &\XSolidBrush   &\CheckmarkBold  &\CheckmarkBold   \\  
{Q}  &\XSolidBrush    &\XSolidBrush      &\XSolidBrush   &\CheckmarkBold &\CheckmarkBold  &\CheckmarkBold   \\
{S}  &\XSolidBrush    &\XSolidBrush      &\CheckmarkBold &\XSolidBrush   &\XSolidBrush    &\CheckmarkBold   \\
{E}  &\XSolidBrush    &\CheckmarkBold    &\XSolidBrush   &\XSolidBrush   &\CheckmarkBold  &\XSolidBrush     \\
{H}  &\XSolidBrush    &\XSolidBrush      &\XSolidBrush   &\XSolidBrush   &\XSolidBrush    &\CheckmarkBold   \\  \bottomrule
\end{tabular}
\end{center}
\end{table}

\begin{table*}
\centering
\setlength{\tabcolsep}{6.5pt}
\caption{AUC on different datasets (higher is better). The number of latent dimensions for all comparison methods (except for POP) is set to $K=10$. For \md, we test different orders of the Markov Chain (i.e.,~1, 2, and 3). On the right we demonstrate the improvement of FPMC vs.~BPR-MF, \md{} vs.~FISM, \md{} vs.~FPMC, and \md{} vs.~the best baselines. } \label{table:auc_k10}
\begin{tabular}{lcccccccccccccccccccc} \toprule
\multirow{2}{*}{Dataset} &(a)    &(b)    &(c)    &(d)    &(e)      &(f-1)  &(f-2)  &(f-3)    &\multicolumn{4}{c}{\% improvement}  \\  \cline{10-13}
                         &POP    &BPR-MF &FISM   &FMC    &FPMC     &\md    &\md    &\md      &e vs.~b  &f vs.~c  &f vs.~e          &f vs.~best\\ \midrule
\emph{Amazon-Office}     &0.6427 &0.6736 &0.7113 &0.6874 &0.6891   &0.7211 &0.7224 &0.7221   & 2.30\%  & 1.56\%  & \textbf{4.83\%} & 1.56\% \\
\emph{Amazon-Auto}       &0.5870 &0.6379 &0.6736 &0.6452 &0.6446   &0.6910 &0.6904 &0.6901   & 1.05\%  & 2.58\%  & \textbf{7.20\%} & 2.58\% \\
\emph{Amazon-Game}       &0.7495 &0.8483 &0.8639 &0.8401 &0.8502   &0.8793 &0.8813 &0.8817   & 0.22\%  & 2.06\%  & \textbf{3.71\%} & 2.06\% \\
\emph{Amazon-Toy}        &0.6240 &0.7020 &0.7499 &0.6665 &0.7061   &0.7625 &0.7645 &0.7652   & 0.58\%  & 2.04\%  & \textbf{8.37\%} & 2.04\% \\
\emph{Amazon-Cell}       &0.6959 &0.7212 &0.7755 &0.7359 &0.7396   &0.7982 &0.8009 &0.8006   & 2.55\%  & 3.27\%  & \textbf{8.29\%} & 3.27\% \\
\emph{Amazon-Clothes}      &0.6189 &0.6513 &0.7085 &0.6673 &0.6672   &0.7255 &0.7256 &0.7259   & 2.44\%  & 2.46\%  & \textbf{8.80\%} & 2.46\% \\
\emph{Amazon-Elec}       &0.7837 &0.7927 &0.8210 &0.7992 &0.7985   &0.8411 &0.8438 &0.8444   & 0.73\%  & 2.85\%  & \textbf{5.75\%} & 2.85\% \\
\emph{Epinions}          &0.4576 &0.5520 &0.5818 &0.5532 &0.5477   &0.6014 &0.6050 &0.6048   &-0.78\%  & 3.99\%  &\textbf{10.46\%} & 3.99\% \\
\emph{Foursquare}        &0.9168 &0.9506 &0.9230 &0.9441 &0.9485   &0.9626 &0.9621 &0.9618   &-0.22\%  & 4.29\%  & \textbf{1.49\%} & 1.26\% \\ 
\midrule
\emph{Avg.} ($K=10$)     &0.6751 &0.7255 &0.7565 &0.7265 &0.7324   &0.7759 &0.7773 &0.7774   & 0.99\%  & 2.79\%  & \textbf{6.54\%} & 2.45\% \\ 
\emph{Avg.} ($K=20$)     &0.6751 &0.7285 &0.7580 &0.7293 &0.7344   &0.7780 &0.7795 &0.7788   & 0.88\%  & 2.83\%  & \textbf{6.44\%} & 2.54\% \\ \bottomrule
\end{tabular}
\end{table*}

These baselines are designed to demonstrate (1) the performance achieved by state-of-the-art sequentially-\emph{unaware} recommendation methods (BPR-MF and FISM) and purely 
sequential methods (MC); (2) the effectiveness of the state-of-the-art sequential prediction method by combining BPR-MF and MC (FPMC); and (3) the strength of our proposed combination of a similarity-based algorithm and (high-order) Markov Chains (\md).

\subsection{Performance and Quantitative Analysis}
For simplicity and fair comparison, the number of dimensions of user/item representations (or the rank of the matrices) in all methods is
fixed to the same number $K$. 
We experimented with different values of $K$ and demonstrate our results in Table \ref{table:auc_k10} ($K=10$). 
Due to the sparsity of these datasets, no algorithm observed significant performance improvements when increasing $K$ beyond 10 (see the last row of Table \ref{table:auc_k10} for the average accuracy of all methods when setting $K$ to 20).

For clarity, on the right of the table we show the percentage improvement of a variety of methods---FPMC vs.~BPR-MF, \md{} vs.~FISM, \md{} vs.~FPMC, and \md{} vs.~the best baselines.
We make a few comparisons and summarize our findings as follows.

\xhdr{FISM vs.~BPR-MF.} 
BPR-MF and FISM are two powerful methods to model users'
personalized preferences, i.e., long-term dynamics. Although ultimately they all factorize a matrix at their core, they 
differ significantly both in terms of the rationales they follow and performance they achieve. BPR-MF relies on factorization of the \emph{user-item} interaction matrix and parameterizes each user with a $K$-dimensional vector. In contrast, FISM is based on the factorization of the \emph{item-item} similarity matrix and relaxes the need to explicitly parameterize users, who may only have a few actions in the training set. According to our experimental results, FISM exhibits
significant improvements over BPR-MF on all datasets (over 4 percent on average), which makes it a strong building-block for our sequential prediction task.

\xhdr{FMC vs.~BPR-MF \& FISM.} 
Compared to BPR-MF and FISM, FMC focuses on capturing sequential patterns among items, i.e., short-term dynamics. Notably, FMC achieved comparable prediction accuracy with BPR-MF. This suggests that sequential patterns are important dynamics and that it would be limiting to only consider long-term user preferences. 

\xhdr{FPMC vs.~BPR-MF \& FMC.} 
BPR-MF and FMC are limited by missing an important type of dynamic prevalent in our datasets. FPMC combines them and emerges as a comprehensive model that is both personalized and sequentially-aware. Quantitatively, FPMC is the strongest among the three---0.99 percent better than BPR-MF and 0.80 percent better than FMC on average (when $K=10$).

\xhdr{\md{} vs.~FPMC.}
By fusing FISM, which is strong at modeling long-term dynamics on sparse data, and Markov Chains, \md{} enhances the performance of FISM by as much as 2.79 percent on average, compared to 0.99 percent achieved by FPMC over BPR-MF. Comparing \md{} with FPMC, we found that (1) \md{} beats FPMC significantly on all datasets, with a large improvement of 6.54 percent on average, and (2) the improvement is even larger on sparse datasets like \emph{Amazon-Auto}, \emph{Amazon-Clothes}, and \emph{Epinions}. The superior performance of \md{} on various datasets demonstrates its efficacy to handle real-world datasets. 

\xhdr{Orders of Markov Chains.}
Generally, the performance of \md{} gets better on most datasets when increasing the order of the Markov Chains involved (from 1 to 3 in our experiments), indicating that earlier actions are also useful for prediction. 
On the other hand, small orders seem to be enough to achieve good performance, presumably because sequential patterns do not involve actions from a long time ago.

\xhdr{\md{} vs.~others.} 
The rightmost column of Table \ref{table:auc_k10} demonstrates the performance improvement of \md{} versus the best baseline method in each case. We found that \md{} outperforms all baselines in all cases with an enhancement of 2.5 percent on average.

\subsection{Reproducibility}
The hyperparameter $\alpha$ in \eq{eq:highorder} is set to 0.2 on all datasets. Regularization hyperparameters are always tuned with grid search using the validation set $\mathcal{V}$. $\lambda_{\Theta}$ (in \eq{eq:sgd}) yielded the best performance when set to 0.1 in most cases. The learning rate $\epsilon$ is set to 0.01.

\subsection{Training Efficiency}
All experiments were performed on a single machine with 8 cores and 64GB main memory. The largest dataset---\emph{Amazon Electronics} takes around 4 hours to train \md{} with third-order Markov Chains (i.e., $L=3$) and 20 latent dimensions (i.e., $K=20$).
It is easy to verify that \eq{eq:highorder} takes $\mathcal{O}(LK)$ for prediction. Since $L$ and $K$ are usually small numbers for sparse datasets, the computational cost is manageable.

\section{A Study on the Effect of Data Sparsity}
We proceed by further studying the effect of dataset sparsity on different methods. To this end, we perform experiments on a popular dataset---\emph{MovieLens-1M},\footnote{http://grouplens.org/datasets/movielens/1m/} which is \emph{dense} and consists of about 1 million ratings from 6,040 users on 3,706 movies dated from April 2000 to February 2003. Again, we convert all star-ratings to implicit feedback for our experiments. The number of dimensions $K$ is set to $10$.

We construct a sequence of datasets each with a different threshold $N$ on the number of recent user actions; that is, a dataset is constructed by taking only the $N$ \emph{most recent} actions of each user. Actions beyond this point are dropped. Note that sampling is not used as it would break the sequential characteristic exploited by the models. We decrease the threshold $N$ from 50 to 5 (leading to a series of increasingly sparse datasets) and observe the performance variation of all methods.
Statistics of the datasets are summarized in Table \ref{table:ml1m}.
Experimental results are collected in Table \ref{table:auc_ml}. As we can see from the table, the accuracy of all methods drops if we decrease the threshold from 50 to 5. This makes sense since we have less information regarding both users and transitions among items. In this section, we compare \md{} with FPMC to answer a series of questions on their effectiveness based on the results in Table \ref{table:auc_ml}.

\begin{table}
\centering
\setlength{\tabcolsep}{3pt}
\caption{Statistics of the \emph{MovieLens} Datasets.}\label{table:ml1m}
\begin{tabular}{lccccccc} \toprule
Dataset    	   &Threshold   &\parbox{0.12\linewidth}{\centering\#users ($|\mathcal{U}|$)}    &\parbox{0.12\linewidth}{\centering\#items ($|\mathcal{I}|$)}   &\#actions   &\parbox{0.12\linewidth}{\centering avg. \#actions /user}  &\parbox{0.12\linewidth}{\centering avg. \#actions /item}  \\ \midrule 
\emph{ML-50}   &50 &6,040       &3,467    &215,676    &35.71  &62.21  \\ 
\emph{ML-30}   &30 &6,040       &3,391    &152,160    &25.19  &44.87  \\  
\emph{ML-20}   &20 &6,040       &3,324    &111,059    &18.39  &33.41  \\ 
\emph{ML-10}   &10 &6,040       &3,114    &59,610     &7.13   &19.14  \\
\emph{ML-5}    &5  &6,040       &2,848    &30,175     &5.00   &10.60  \\  \bottomrule
\end{tabular}
\end{table}

\begin{table}
\centering
\setlength{\tabcolsep}{2pt}
\caption{AUC on different datasets (higher is better). The number of latent dimensions $K$ is set to $10$. For \md, we test different orders of the Markov Chain (i.e.,~1, 2, and 3). } 
\begin{tabular}{lcccccccccccc} \toprule
\multirow{2}{*}{Dataset} &(a)    &(b)    &(c)    &(d)    &(e)      &(f-1)  &(f-2)  &(f-3)    \\  
                         &POP    &BPR-MF &FISM   &FMC    &FPMC     &\md    &\md    &\md      \\ \midrule
\emph{ML-50}             &0.8032 &0.8587 &0.8564 &0.8566 &0.8825   &0.8802 &0.8837 &0.8865   \\
\emph{ML-30}             &0.7980 &0.8523 &0.8515 &0.8463 &0.8674   &0.8748 &0.8794 &0.8797   \\
\emph{ML-20}             &0.7919 &0.8447 &0.8476 &0.8357 &0.8503   &0.8704 &0.8735 &0.8728   \\
\emph{ML-10}             &0.7722 &0.8728 &0.8276 &0.8026 &0.8301   &0.8540 &0.8546 &0.8551   \\
\emph{ML-5}              &0.7352 &0.7551 &0.7711 &0.7275 &0.7458   &0.7945 &0.7940 &0.7931   \\  \bottomrule
\end{tabular}
\label{table:auc_ml}
\end{table}

\subsection{How Much do Short-term Dynamics Help?}
\begin{figure}[!t]
\centering
\includegraphics[width=0.9\linewidth]{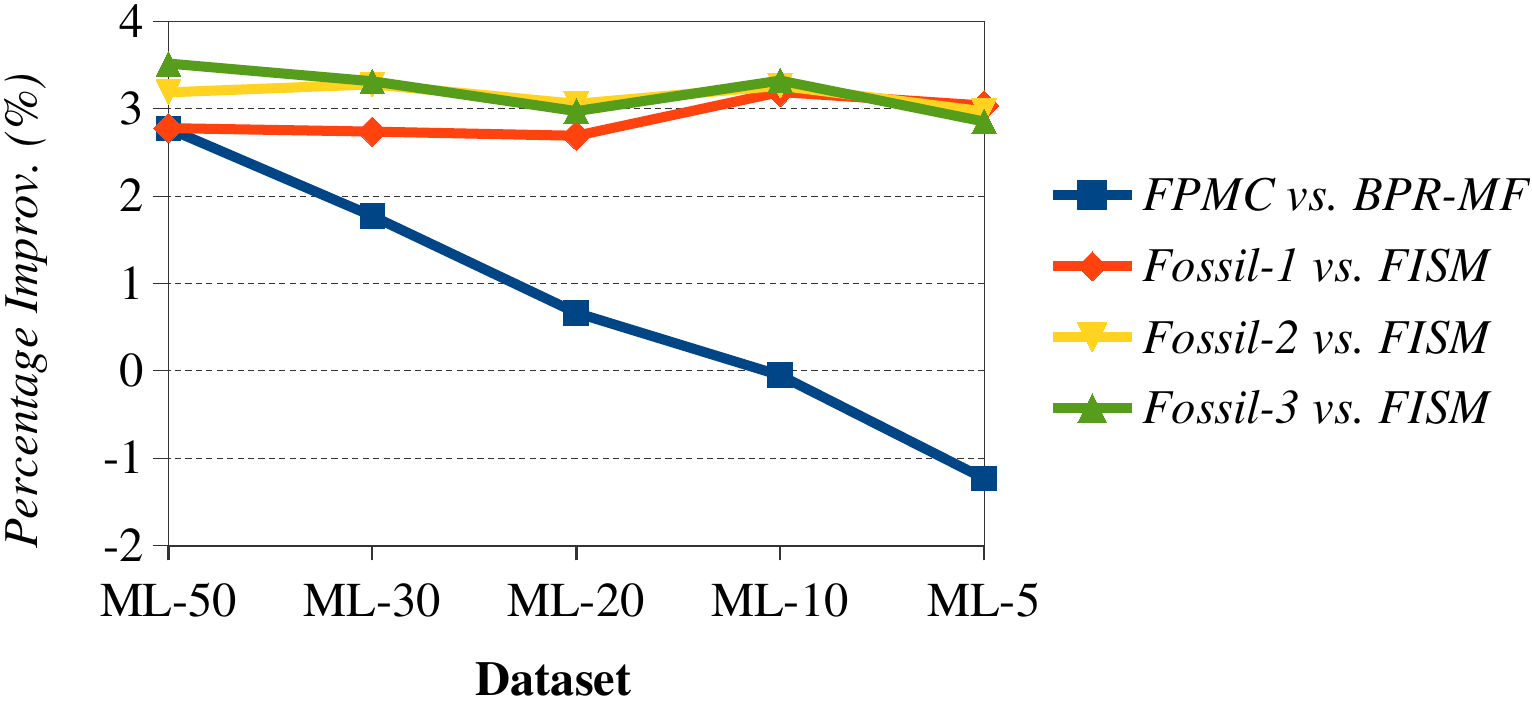}
\caption{
Improvement from modeling short-term dynamics. Such dynamics (with 1st-, 2nd-, and 3rd-order Markov Chains) are {consistently} helpful for FISM, in contrast to the deteriorating improvement of FPMC over BPR-MF when the dataset is sparse.}
\label{fig:vs}
\end{figure}

We begin by investigating the improvements due to modeling sequential patterns on top of BPR-MF and FISM.
In Figure \ref{fig:vs} we demonstrate the improvement of FPMC over BPR-MF, in contrast to 
that achieved by \md{} (with 1st-, 2nd-, and 3rd-order Markov Chains) over FISM. From the figure we observe that the improvement of FPMC over BPR-MF decreases as data become sparser, due to the amount of additional parameters introduced to model short-term dynamics. In contrast, \md{} introduces only a small amount of parameters (i.e., the weighting vectors $\eta$ and $\eta^u$), and thus outperforms FISM more consistently. Additionally, high orders of Markov Chains appear to help more when the dataset is dense, e.g., \emph{ML-50} vs.~others.

\begin{figure}[!t]
\centering
\includegraphics[width=0.9\linewidth]{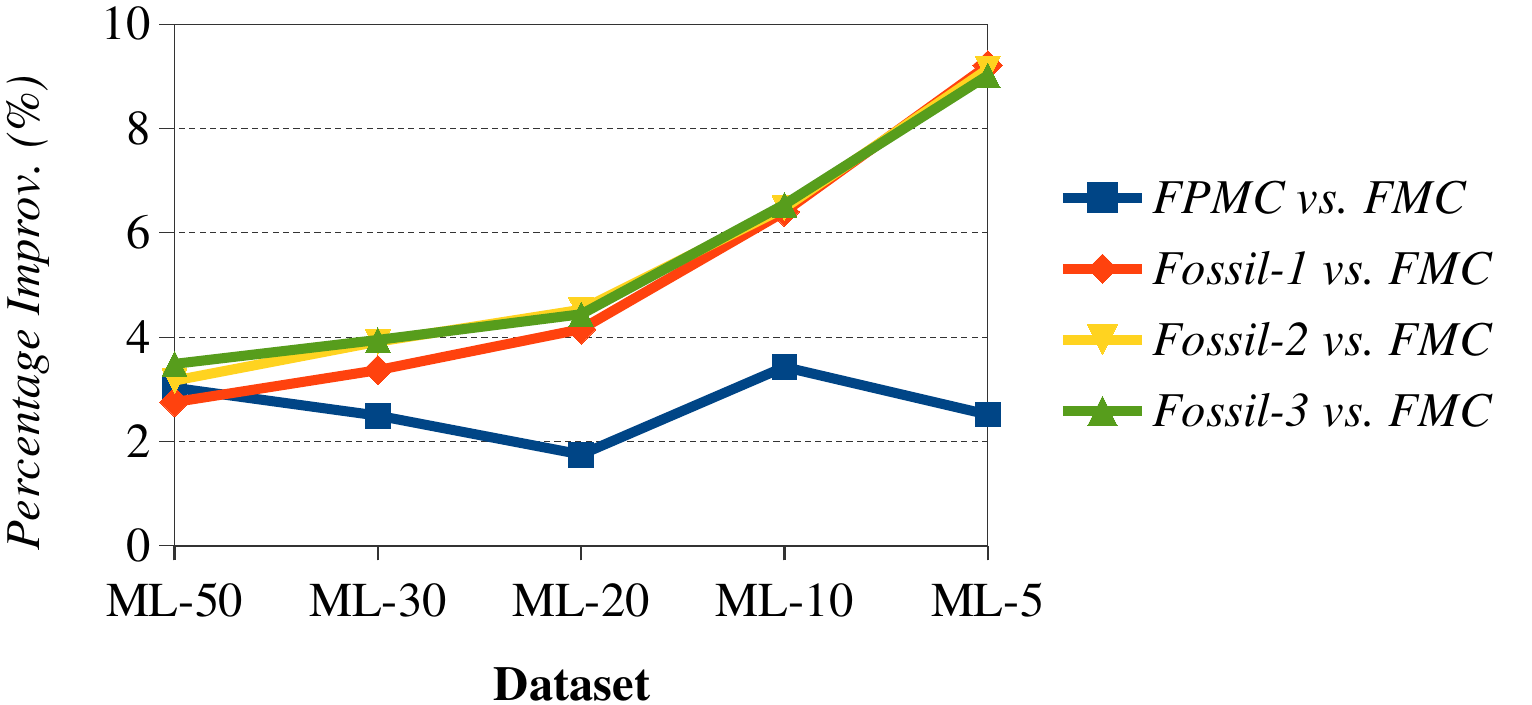}
\caption{
Improvement from modeling long-term dynamics on top of Markov Chains. The improvement achieved by \md{} increases as datasets get sparser.}
\label{fig:vsFMC}
\end{figure}

\begin{figure}[!t]
\centering
\includegraphics[width=0.9\linewidth]{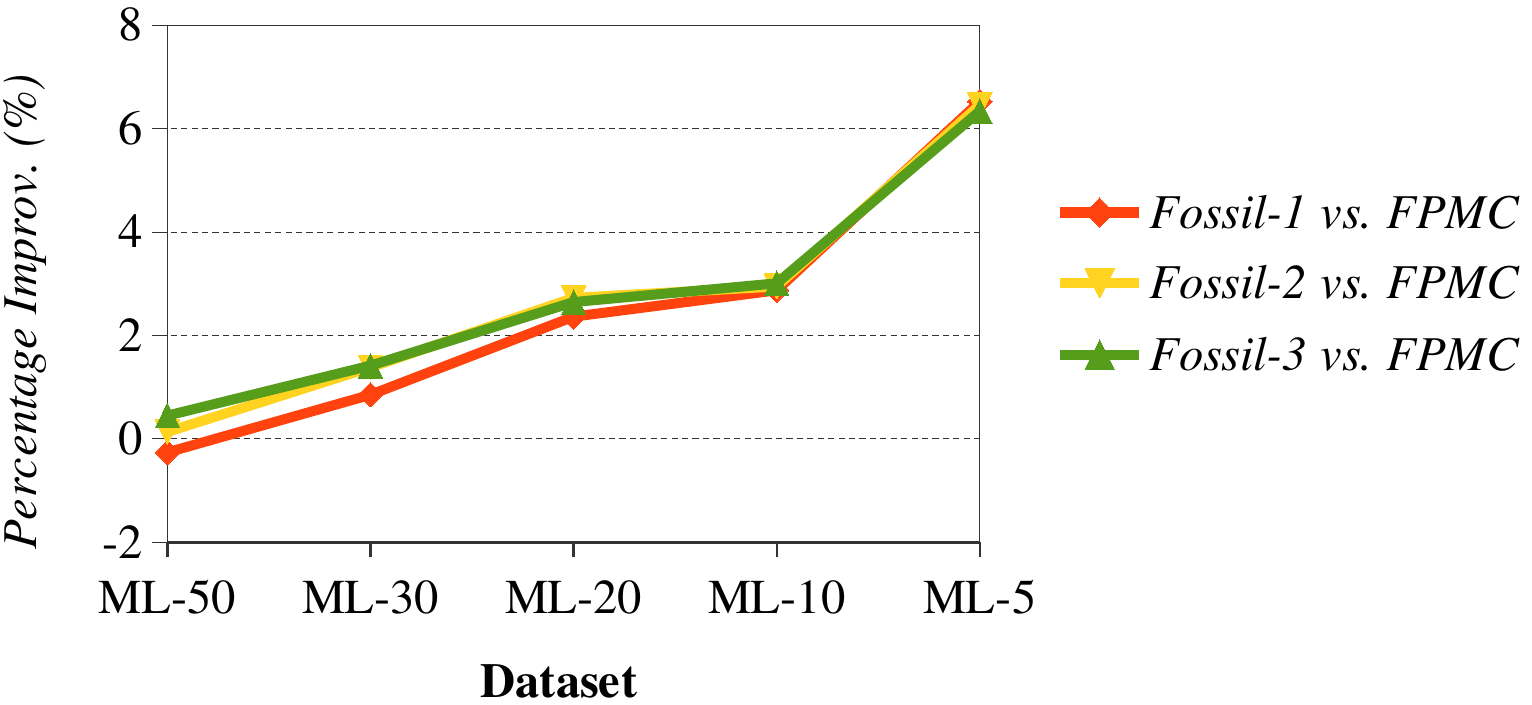}
\caption{
Improvement achieved by \md{} over FPMC 
as dataset sparsity increases.
}
\label{fig:vsFPMC}
\end{figure}

\subsection{How Much do Long-term Dynamics Help?}
One may consider FPMC and \md{} as two different methods to enhance Markov Chains by incorporating long-term user preferences. To demonstrate the benefits of modeling long-term dynamics, in Figure \ref{fig:vsFMC} we show the amount of improvement achieved by FPMC and \md{} over FMC.
As datasets become sparser, the performance gap between \md{} and FMC increases 
by as much as 9 percent, in contrast to the relatively `flat' improvement 
of around 3 percent from FPMC. This demonstrates the strength of \md{} as well as the compatibility of FISM and Markov Chains since they are both ultimately modeling relationships among items.

\subsection{\emph{\md{}} and FPMC}
In Figure \ref{fig:vsFPMC} we demonstrate the performance gain of our proposed method over FPMC. \md{} outperforms FPMC increasingly as sparsity grows. Note that FPMC requires as many as 50 actions per user in order to achieve comparable performance with \md. 

For further investigation, we also performed additional experiments where we reduced the number of parameters of FPMC by setting the
matrix $\mathbf{M} = \mathbf{N}$ in \eq{eq:fpmc}, 
in the hope that it could favor sparse datasets. However, no significant improvement (or even worse prediction accuracy) was observed. 

To sum up, the two key components of \md{}, i.e., the similarity-based method and (high-order) Markov Chains, both contribute considerably to its performance especially on sparse datasets. And thus the precise combination of the two generates strong results for the sparse sequential recommendation task.

\section{Visualization and Qualitative Analysis}
In this section, we visualize the learned \md{} model and qualitatively analyze our findings. We choose to visualize the results achieved on \emph{Clothing, Shoes and Jewelry} dataset from \emph{Amazon.com} (see Section \ref{sec:dataset}) due to its large size, significant variability, and the convenience to demonstrate user actions. The model we use for visualization is the first-order \md{} model trained on the dataset with $K$ set to $10$.

\begin{figure}[!t]
\centering
\includegraphics[width=\linewidth]{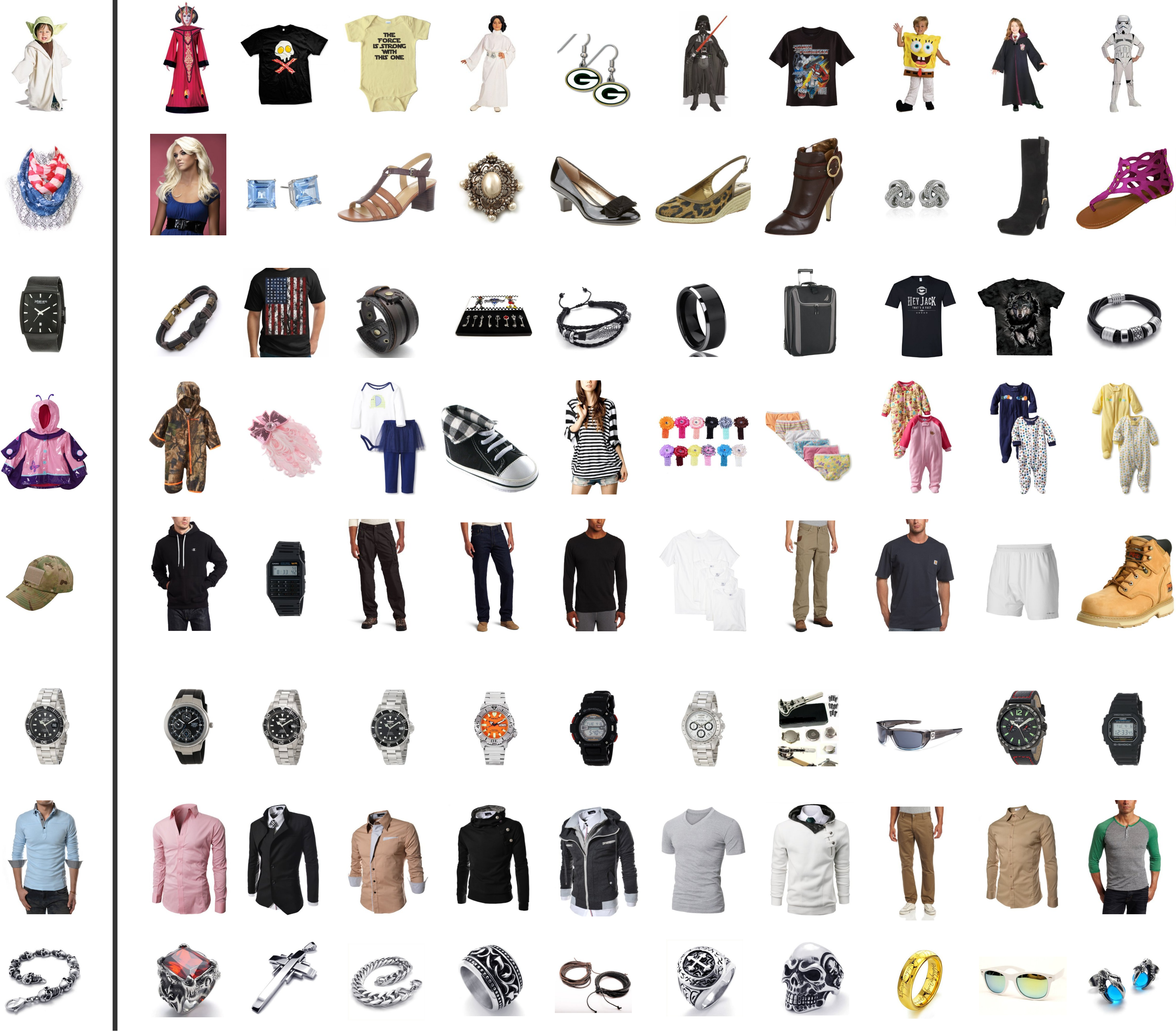}
\caption{Demonstration of item transitions. On the left are a few sampled items (or queries) from the \emph{Clothing, Shoes and Jewelry} dataset. On the right are the top-ranked items (from the same dataset) that each query is likely to transition to, according to $\argmin_{j \in \mathcal{I}} \langle \mathbf{P}_{query}, \mathbf{Q}_j \rangle$.}
\label{fig:query}
\end{figure}

\subsection{Visualizing Sequential Dynamics}
First, we visualize the transition among items to answer questions like `What kind of outfits are compatible with this outdoor cap?'. \md{} encodes this dynamic by the inner product of $\mathbf{P}_i$ and $\mathbf{Q}_j$ where $i$ is the item already consumed and $j$ the item considered for recommendation. Quantitatively, given a `query' item $i$ at the current time step, items that are most likely to appear at next step are computed according to  
\begin{equation} \label{eq:query}
\argmin_{j \in \mathcal{I}}~\langle \mathbf{P}_i, \mathbf{Q}_j \rangle.
\end{equation}

For demonstration, we take a few samples from the above dataset, as shown on the left of the separator in Figure \ref{fig:query}. Next, we use them as queries to get corresponding recommendations according to \eq{eq:query}. Items retrieved for each query are shown on the right. We make two observations from this figure. On the one hand, 
although the model is \emph{unaware} of the identity of items, it learns the underlying homogeneity correctly, as we see from the first row (i.e., the Star Wars theme) and last three rows (i.e., watches, shirts and jewelry respectively). On the other hand, items from different subcategories are surfaced to generate compatible outfits, e.g.~rows 2 and 5. 

\begin{figure} 
\centering
\includegraphics[width=\linewidth]{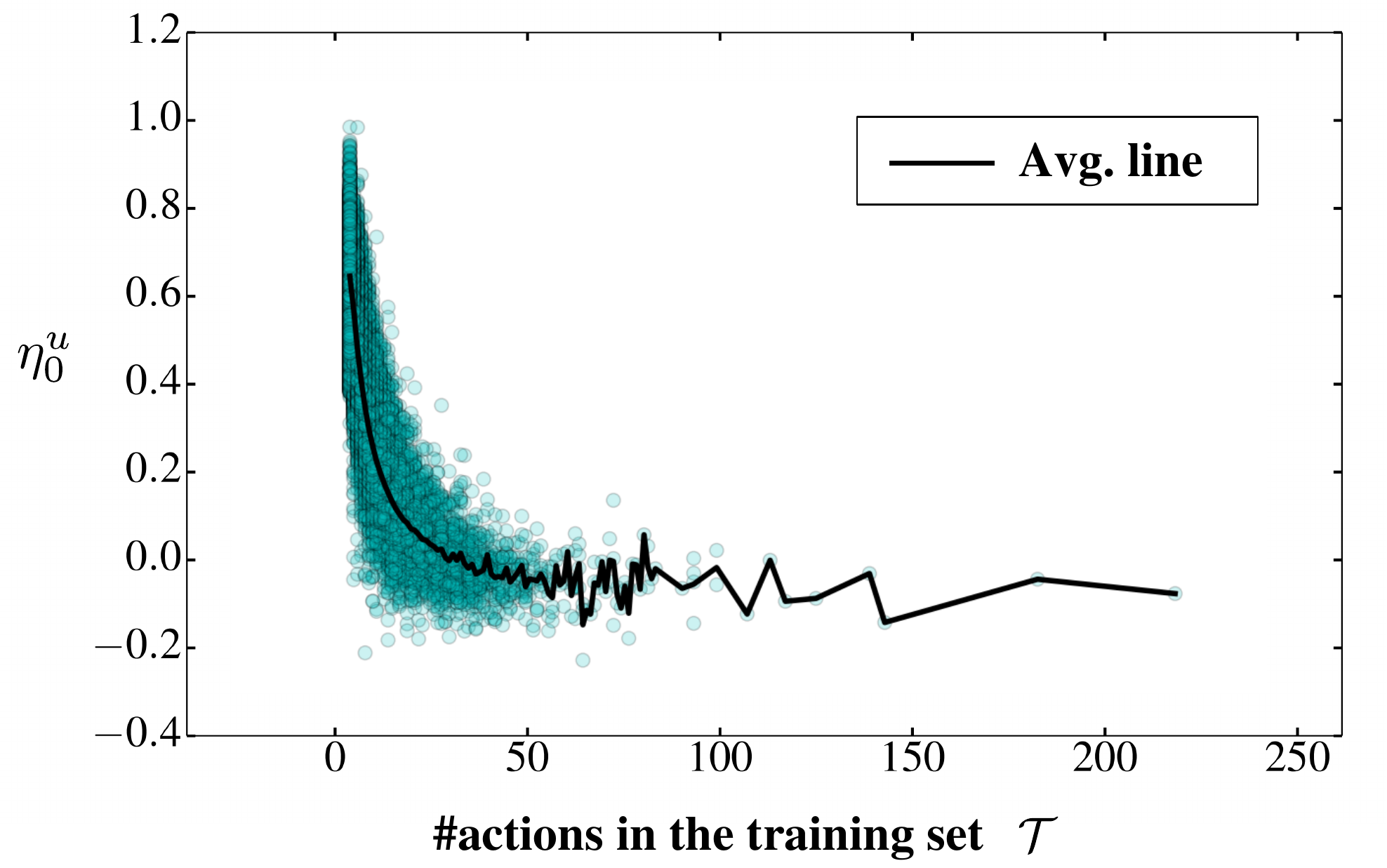}
\caption{Demonstration of the importance of short-term dynamics for different users. \md{} learns to assign higher weights to cold users to rely more on sequential patterns.}
\label{fig:etaU}
\end{figure}

\subsection{Visualizing Personal Dynamics}
Recall that each user $u$ is parameterized with a vector $\eta^u$ to model the personalized comparative importance of short-term dynamics versus long-term dynamics. The scatter plot in Figure \ref{fig:etaU} shows such weights ($\eta^u_0$) learned for users in the dataset. 

From the scatter plot we make a few observations as follows. (1) \md{} gives `cool' users (users with few actions) higher weights, which makes sense since sequential patterns have to carry more weight when we do not have enough observations to infer users' preferences. This also confirms that it is necessary to model short-term dynamics on sparse datasets in order to enhance the performance of item recommendation methods like FISM. (2) As we increment the number of actions, \md{} relies more on long-term preferences as they become increasingly accurate.

\begin{figure} 
\centering
\includegraphics[width=\linewidth]{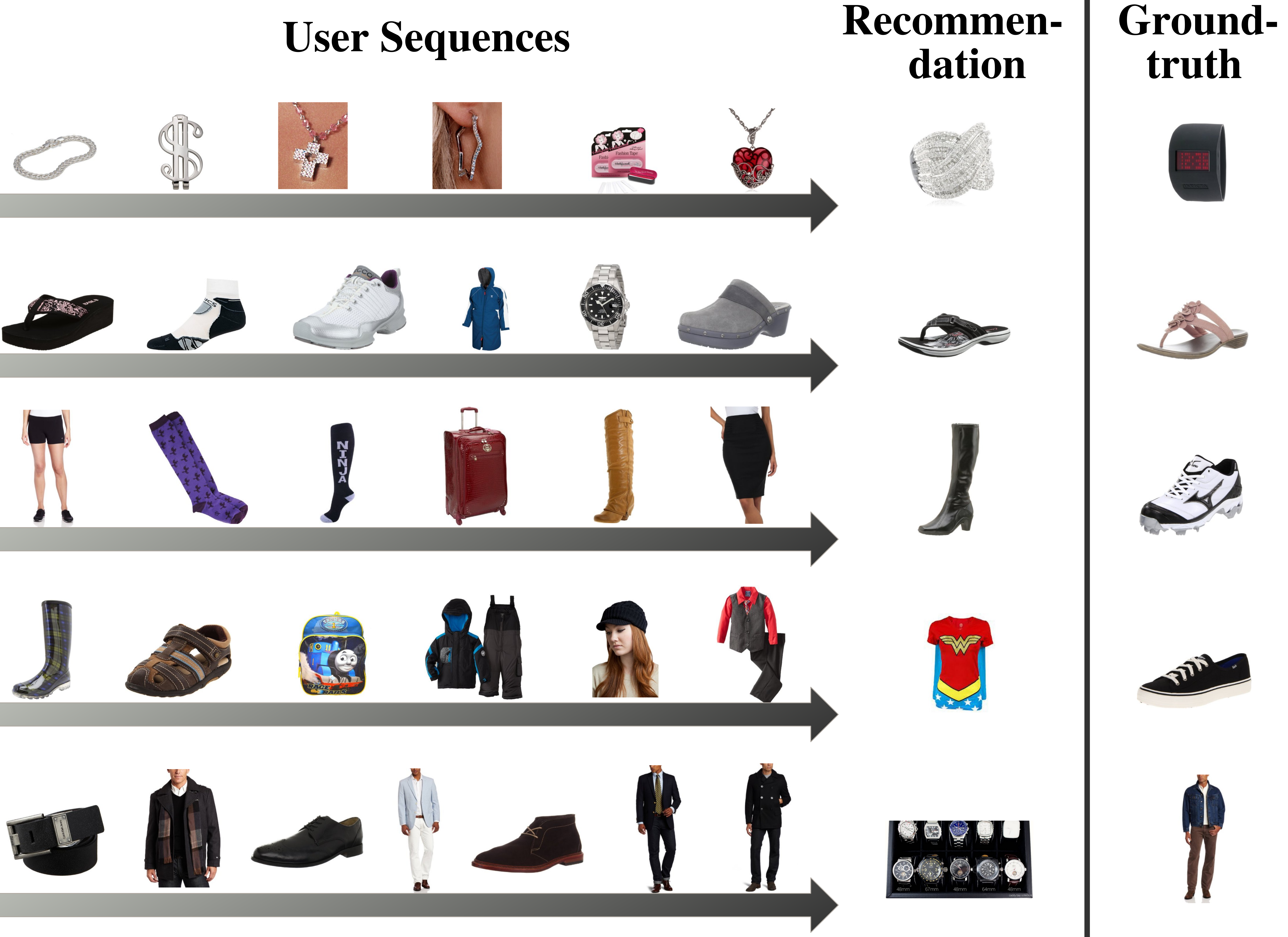}
\caption{Demonstration of recommendations made for users with large $\eta_u^0$ values, indicating strong `sequential consistency.'}
\label{fig:recomm}
\end{figure}

In Figure \ref{fig:recomm} we demonstrate recommendations made for a few users as well as the corresponding ground-truth items (from the test set $\mathcal{E}$). Here the users are sampled from those with the largest $\eta^u_0$ and at least 5 actions in the training set $\mathcal{T}$. The threshold is used so that $\eta^u_0$ is forced to capture the `sequential consistency' of user $u$ to some degree, instead of merely the user sparsity involved. From Figure \ref{fig:recomm} we can observe a certain amount of such consistency within each sequence. E.g., jewelry (row 1), wearables for boys (row 4) and business men (row 5). 

In conclusion, it is precisely the ability to carefully accommodate multiple types of dynamics as well as personalization that makes \md{} a successful method to handle the sequential recommendation task.

\section{Conclusion}
In this paper, we proposed a new method, \md{}, that fuses similarity-based models with Markov Chains to predict personalized sequential behavior. 
We performed extensive experiments on multiple large, real-world datasets, and found that \md{} outperforms existing methods considerably. 
We studied the effect of sparsity on different methods and found that \md{} is especially strong when the prediction task is challenged by sparsity issues.
We visualized the learned \md{} model on a large dataset from \emph{Amazon.com} and observed that it captures sequential and personalized dynamics in a reasonable way, along with the favorable quantitative results achieved.

\ \\
\small
\xhdr{Acknowledgments.} This work is supported by NSF-IIS-1636879, and donations from Adobe, Symantec, and NVIDIA.

\balance
\setlength{\bibsep}{0pt}
\small
\bibliographystyle{IEEEtran}
\bibliography{IEEEabrv}

\end{document}